\documentclass[12p]{article}
\usepackage{latexsym}
\usepackage{color}
\usepackage{amsmath}
\usepackage{epsfig}
\usepackage{hyperref}
 \usepackage{dcolumn}
\newcommand{\ortala}[1]{\begin{center}#1\end{center}}

\newcommand{\ket}[1]{\left|#1\right\rangle}

\newcommand{\sand}[3]{\left\langle #1\left|#2\right|#3\right\rangle}
\newcommand{\sandd}[1]{\left\langle #1\right\rangle}

\newcommand{\integ}[3]{{{\underset{#1 }{\overset{#2}{\displaystyle\int}}}#3}}

\newcommand{\summ}[3]{{{\underset{#1 }{\overset{#2}{\displaystyle\sum}}}#3}}

\newcommand{\re}[1]{(\ref{#1})}

\newcommand{\eq}[2]{\begin{equation}\label{#1}  #2\end{equation}}

\newcommand{\paran}[1]{\left(#1\right)}

\newcommand{\sch}[1]{Schrodinger}
\newcommand{\ktur}[2]{\frac{\partial #1}{\partial #2}}

\usepackage{amssymb}
\usepackage{latexsym}
\begin{document}

\ortala{\large\textbf{Magnetocaloric Properties of the Ising nanotube}}

\ortala{\textbf{\"Umit Ak\i nc\i \footnote{\textbf{umit.akinci@deu.edu.tr}}}}

\ortala{\textit{Department of Physics, Dokuz Eyl\"ul University,
TR-35160 Izmir, Turkey}}

\section{Abstract}

The magneticaloric properties of the Ising nanotube constituted by arbitrary core spin 
values $S_c$ and the shell spin values $S_s$ have been investigated by 
mean field approximation. During this investigation, several quantities 
have been calculated, such as isothermal magnetic entropy change, full width at half  maximum value and the refrigerant capacity.  
The variation of these quantities with the values of the spins and exchange interaction
between the core and shell is determined. Besides, recently experimentally observed  double peak behavior in the variation of the isothermal magnetic entropy change with the temperature is obtained for the nanotube.

\section{Introduction}\label{introduction}

Magnetocaloric effect (MCE) is defined as an occurred temperature change in the material when it is subject to the magnetic field. It was first observed in Iron \cite{ref1} and theoretically explained after many years 
\cite{ref2,ref3}. MCE is simply based on the variation in different contributions to the entropy. The entropy of a magnetic material is composed of three independent parts namely, the electronic part, lattice part, and magnetic part.  
Under adiabatic changes, the total entropy of the material is constant. This means that, occurred a change in one part of the entropy should be balanced by other parts. Then if one increases the magnetic part of the entropy by an adiabatic process, the lattice part should decrease (by an assumption of the constant electronic part of the entropy). Decreasing lattice entropy manifests itself as a reduction in the temperature of the material. In this way, 
one can construct a thermodynamical cycle, in which at one step the material is at the temperature $T_1$ and at another step it has the temperature $T_2>T_1$.

Refrigerant capacity (RC) is the amount of heat
that can be transferred from the cold end (at temperature $T_1$)  to
the hot end (at temperature $T_2$ ) in one thermodynamical cycle. This quantity is one of the quantities which measure
the suitability of the magnetic material for magnetocaloric purposes. It is in relation to another quantity namely 
 isothermal magnetic entropy change (IMEC). In order to obtain a large adiabatic temperature change, the material should have a large IMEC, 
and a large RC. On the other hand a good candidate has sufficient thermal conductivity for the aim of easy heat exchange. 

The typical behavior of the IMEC by the temperature includes peak at a critical temperature of the material. 
Generally, bulk magnetocaloric materials display larger  IMEC peaks
but with negligible or very low RC values. 
On the other side, nanosystems show reduced IMEC values. But their IMEC curve spread over a wide temperature range
and this fact sometimes yields larger RC (in comparison to the bulk counterparts). Thus  
they  are promising candidates for magnetic
refrigeration \cite{ref4,ref5}. For instance, it has been shown that the geometrical confinement
of $Dy$ and $Ho$ can lead to an enhanced magnetocaloric effect
in comparison to the bulk counterparts \cite{ref6,ref7}.
Similarly, it has been shown for $La_{0.7} Sr_{0.3} MnO_3$ nanotube arrays, 
the bulk sample exhibits higher IMEC but 
nanotubes present an expanded temperature dependence
of IMEC  curves that spread over a broad temperature range \cite{ref8,ref9}. 

As explained in Sec. \ref{formulation}, IMEC is related to the magnetization change with the temperature.   
If the magnetization rapidly changes over some interval of the temperature, it is said that large MCE obtained. 
Nanotubes are promising materials for obtaining efficient MCE.  
For instance, large MCE, associated with the sharp change in magnetization of the
$Gd_2 O_3$ nanotubes has been shown experimentally \cite{ref10,ref11}. Another example of experimental 
MCE in nanotubes is the structural defect-induced MCE in
$Ni_{0.3} Zn_{0.7} Fe_2 O_4$ graphene (NZF/G) nanocomposites \cite{ref12}. 

As seen in these examples, experimental studies are up to date for MCE in nanotubes. 
Although, MCE in nano systems is an active research area  for experimentalists, 
to the best of our knowledge MCE on nanotube geometry has not been worked out, theoretically.
But, from the theoretical side both of the magnetic behavior of these systems well studied. 
After the first theoretical
treatments of the Ising model on nanotube geometry \cite{ref13} by effective field approximation,
the first results for the anisotropic Heisenberg model on nanotube geometry have been obtained within the same methodology \cite{ref14}. As studied in this work in terms of the MCE, mixed spin models have been worked out
for obtaining the magnetic properties. The magnetic properties of the 
spin (1/2-1)  mixed system on nanotube geometry has been worked out within the improved mean-field approximation \cite{ref15} and Monte Carlo simulation \cite{ref16,ref17,ref18}. Also hysteresis and magnetic properties of the spin-1/2 spin-1 nanowire have been determined by Monte Carlo simulations \cite{ref19}.
The magnetic and hysteresis behaviors of the higher spin models are also well known theoretically. 

The magnetic properties of the spin-1 and spin 3/2 nanotube has been determined within the Monte Carlo simulation  
\cite{ref20} and quantum simulation treatment \cite{ref21}. The same model on the nanowire geometry has been investigated by 
Monte Carlo simulation   \cite{ref22}. Spin (1/2-3/2) model on nanotube geometry has been investigated 
within the effective field theory \cite{ref23}
and on a nanowire geometry by Monte Carlo simulation \cite{ref24,ref25}. 
The magnetic phase transition characteristics and hysteresis behaviors of 
spin-3/2 spin -5/2 model on Ising nanowire have been determined by the Monte Carlo simulation \cite{ref26,ref27}.
Besides, hysteresis and compensation behaviors of mixed spin-2 and spin-1
hexagonal Ising nanowire have been studied within the Monte Carlo simulation \cite{ref28}.

The aim of this work is to determine the MCE characteristics of the magnetic nanotube,
by solving the Ising model with several different spin values. 
For this aim, the paper is organized as follows: In Sec. 
\ref{formulation} we
briefly present the model and formulation. The results and
discussions are presented in Sec. \ref{results}, and finally Sec.
\ref{conclusion} contains our conclusions.

\section{Model and Formulation}\label{formulation}

\begin{figure}[h]\begin{center}
\epsfig{file=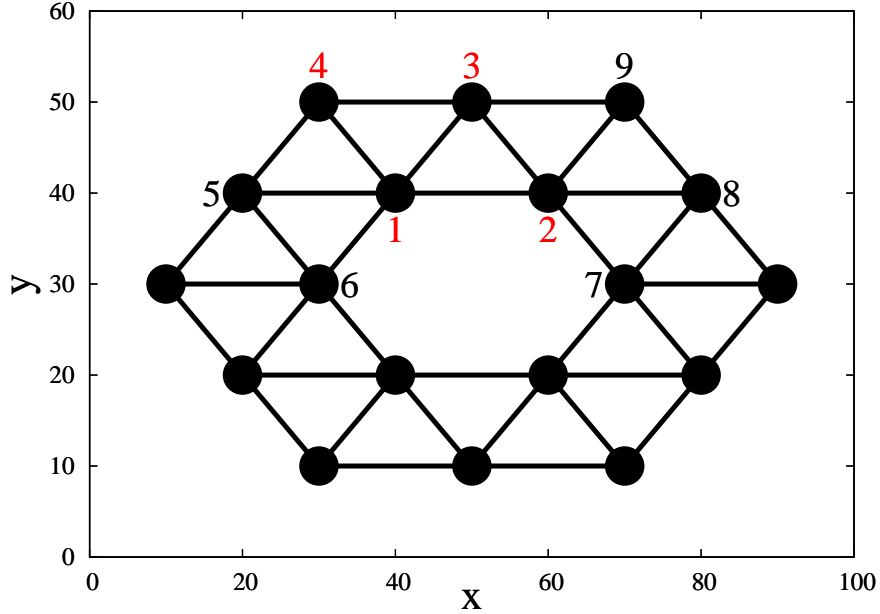, width=12cm}
\end{center}
\caption{Schematic representation of one layer of the nanotube in $xy$ plane. The system periodically extends in $z$ direction. } \label{sek1}
\end{figure}

We can see the schematic representation of the one layer of the nanotube in Fig. \re{sek1}.  As seen in 
Fig. \re{sek1}, one layer of the nanotube consists of two hexagons which is called core (inner hexagon) and 
the shell (outer hexagon). Let the core spins have value $S_c$ and shell spins have $S_s$. We can write 
the Hamiltonian of this system as

\eq{denk1}{
\mathcal{H}=-J_c\summ{<i,j>}{}{\paran{S_i^cS_j^c}}
-J_s\summ{<i,j>}{}{\paran{S_i^sS_j^s}}
-J_{cs}\summ{<i,j>}{}{\paran{S_i^cS_j^s}}
-H\summ{i}{}{S_i}
}
where $S_i^c,S_i^s$ denote the $z$ component of the Pauli spin 
operator at a site $i$ which belongs to the core (c) and shell (s), respectively. 
$J_c$ is the exchange interaction between the nearest neighbor core spins, 
$J_s$ is the exchange interaction between the nearest neighbor shell spins, and 
$J_{cs}$ is the exchange interaction between the nearest neighbor core and shell spins. The former
three sums in Eq. \re{denk1} are taken over the nearest neighbor sites, while last summation is taken over 
all the lattice sites. In Eq. \re{denk1}, $H$ is the longitudinal magnetic field.

In order to include the effect of all exchange interactions, we take four spin cluster. We can construct one 
layer of the nanotube by repetition of this selected cluster. The nanotube consists of repeating layers 
(seen in Fig. \re{sek1}) in $z$ direction.
The Hamiltonian of this 
cluster (which consists of red colored spins in Fig. \re{sek1}) is 
\eq{denk2}{
\mathcal{H}^{(4)}=-J_c\paran{S_1S_2}
-J_s\paran{S_3S_4}
-J_{cs}\paran{S_1S_3+S_1S_4+S_2S_3}
}
$$
-H\summ{i=1}{4}{}S_i
-\summ{i=1}{4}{}h_iS_i. 
$$
Here, $h_i$ are the local fields that represent the interaction of the $i^{th}$ spin with nearest neighbor spins 
that belong to outside of the cluster. These local fields represent the following spin-spin interactions:

\eq{denk3}{
\begin{array}{lcl}
 h_1&=&J_c\paran{S_6+S_{11}+S_{12}}+J_{cs}S_5\\
h_2&=&J_c\paran{S_7+S_{21}+S_{22}}+J_{cs}\paran{S_8+S_9}\\
h_3&=&J_s\paran{S_9+S_{31}+S_{32}}\\
h_4&=&J_s\paran{S_5+S_{41}+S_{42}}.
\end{array}
}
Here the spins which are denoted as $S_{ij}$, where $i=1,2,3,4$ and $j=1,2$ represent the neighbor spins of the spin denoted as $S_i$, which are in the upper and lower plane in $z$ direction. 
The thermal average of  the quantity $\Omega$ can be calculated via 
the exact generalized
Callen-Suzuki identity \cite{ref29}
\eq{denk4}{
\sandd{\Omega}=\sandd{\frac{Tr_4 \Omega \exp{\paran{-\beta 
\mathcal{H}^{(4)}}}}{Tr_4 \exp{\paran{-\beta \mathcal{H}^{(4)}}}}}
} In Eq.  \re{denk4}  $Tr_4$ stands for the partial trace over all the lattice 
sites which belong to the selected cluster and $\beta=1/(kT)$ where $k$ is the 
Boltzmann constant and $T$ is the temperature. 

Let us denote the basis set for this finite cluster by $\{\ket{\phi_i}\}=\ket{s_1s_2s_3s_4}$, where 
$s_k $ is just one spin eigenvalue of the operator $S_k$  
($k=1,2,3,4$). In this representation of the basis set, operators in the 
$4$-spin cluster acts on the bases via

\eq{denk5}{
S_k\ket{s_1s_2s_3s_4}=s_k\ket{s_1s_2s_3s_4} ,
} where $k=1,2,3,4$. Note that, since the system consist of spin-$S_c$ core and spin-$S_s$ shell particles, 
number of bases equals to $\paran{2S_c+1}^2\paran{2S_s+1}^2$. 

Indeed calculation of Eq. \re{denk4} is trivial, since the matrix $\mathcal{H}^{(4)}$ is diagonal for the Hamiltonian 
given in Eq. \re{denk2}, in the chosen basis set.  The diagonal element related to the base $\ket{s_1s_2s_3s_4}$ (which can be obtained 
by applying operator Eq. \re{denk2} to bases according to Eq. \re{denk5}) is given by

\eq{denk6}{
\sand{\phi_i}{\mathcal{H}^{(4)}}{\phi_i}=-J_c\paran{s_1s_2}
-J_s\paran{s_3s_4}
-J_{cs}\paran{s_1s_3+s_1s_4+s_2s_3}
}
$$
-H\summ{i=1}{4}{}s_i
-\summ{i=1}{4}{}h_is_i. 
$$

Let us denote this element as $H^{(4)}\paran{s_1,s_2,s_3,s_4}$, then we can write Eq.  \re{denk4} as 

\eq{denk7}{
\sandd{S_k}=\sandd{\frac{\summ{\{s_1,s_2,s_3,s_4\}}{}{} s_k \exp{\paran{-\beta 
H^{(4)}\paran{s_1,s_2,s_3,s_4}}}}{\summ{\{s_1,s_2,s_3,s_4\}}{}{} \exp{\paran{-\beta 
H^{(4)}\paran{s_1,s_2,s_3,s_4}}}}}, k=1,2,3,4. 
} The summations are taken over all the possible configurations of $(s_1,s_2,s_3,s_4)$.

The core ($m_c$), shell ($m_s$) and total ($m$) magnetizations can be calculated via

\eq{denk8}{
m_c=\frac{1}{2}\paran{\sandd{S_1}+\sandd{S_2}}, \quad 
m_s=\frac{1}{2}\paran{\sandd{S_3}+\sandd{S_4}}, \quad m=\frac{1}{3}\paran{m_c+2m_s}.
}

Up to this point, all equations are exact. But how can local fields in Eq. \re{denk6} be treated? Since our aim is to obtain some general qualitative results about 
the MCE in  nanotube system, it is enough to treat these local fields in a level of mean field, i.e. by 
writing operators in Eq. \re{denk3} as their thermal averages, 

\eq{denk9}{
\begin{array}{lcl}
h_1&=&J_c\paran{2m_1+m_2}+J_{cs}m_3\\
h_2&=&J_c\paran{m_1+2m_2}+J_{cs}\paran{m_3+m_4}\\
h_3&=&J_s\paran{2m_3+m_4}\\
h_4&=&J_s\paran{m_3+2m_4}.
\end{array}
} Note that, the periodicity of the lattice has been used for obtaining the expressions of local fields given in 
Eq. \re{denk9} from Eq. \re{denk3}. In other words, 
\eq{denk10}{
\begin{array}{rcr}
\sandd{S_{11}}=\sandd{S_{12}}=\sandd{S_7}&=&m_1\\
\sandd{S_{21}}=\sandd{S_{22}}=\sandd{S_6}&=&m_2\\
\sandd{S_{31}}=\sandd{S_{32}}=\sandd{S_5}=\sandd{S_8}&=&m_3\\
\sandd{S_{41}}=\sandd{S_{42}}=\sandd{S_9}&=&m_4.
\end{array}
}

By using this approximation, Eq. \re{denk7} gets the form 

\eq{denk11}{
m_k=\frac{\summ{\{s_1,s_2,s_3,s_4\}}{}{} s_k \exp{\paran{-\beta 
H^{(4)}\paran{s_1,s_2,s_3,s_4}}}}{\summ{\{s_1,s_2,s_3,s_4\}}{}{} \exp{\paran{-\beta 
H^{(4)}\paran{s_1,s_2,s_3,s_4}}}}, k=1,2,3,4, 
} where the definitions of local fields given by Eq. \re{denk9} have been used in matrix elements
$H^{(4)}\paran{s_1,s_2,s_3,s_4}$. Then, the magnetizations $m_1,m_2,m_3,m_4$ can be found by numerical
solution of the nonlinear equation system given by Eq. \re{denk11}. Core, shell and the total magnetization can 
be obtained by using Eq. \re{denk8}. Note that, the formulation used here is a generalization of the traditional mean-field to a larger cluster. The effect of using larger clusters can be found in Ref. \cite{ref30}.

In  order to determine the magnetocaloric properties of the system, we calculate the 
isothermal magnetic entropy change (IMEC) when maximum applied  longitudinal field is $H_{max}$, which is given by \cite{ref31}
\eq{denk12}{
\Delta S_M=\integ{0}{H_{max}}{}\paran{\ktur{m}{T}}_H dH.
} 
The other quantitiy of interest is the refrigerant capacity which is defined by  \cite{ref32}
\eq{denk13}{
q=-\integ{T_1}{T_2}{} \Delta S_M\paran{T}_H dT. 
} 
 Here 
$T_1$ and $T_2$ are chosen as those temperatures at which
the magnetic entropy change gains the half of the peak value and this is called as the full width at half 
maximum value (FWHM) of 
the IMEC. This is  
also an important quantity of the MCE.

\section{Results and Discussion}\label{results}

We want to focus on the magnetocaloric properties of the system. The Hamiltonian 
of the system includes four parameters, as one can see from Eq. \re{denk1}. In order to make it possible
for investigation, we have to reduce this number of parameters. For this aim let us choose $J_c=J_s=J$. 
By this unit of energy $J$ (which is positive) we can work with scaled  quantities as
\eq{denk14}{
r=\frac{J_{cs}}{J}, \quad h=\frac{H}{J}, \quad t=\frac{k_BT}{J}.  
} Note that, $h_{max}=1.0$ is chosen in the calculations.  

First, we want to elaborate on IMEC behavior for differently structured nanotubes. For this aim, we depict the 
variation of the IMEC with the temperature for several nanotubes constituted by core spin value 
$S_c=1/2$  in Fig. \re{sek2}
and $S_c=7/2$  in Fig. \re{sek3}. Each figure contains different shell spin values ($S_s$) and core-shell
exchange interaction values ($r$), which are shown in the related figure. We can see from Fig. \re{sek2} that, 
when the spin value of the shell increases, the maximum value of the IMEC occurs in higher temperatures, and the peak value (i.e. height of the peak) of the IMEC decreases. At the same time, the curve gets wider, i.e. FWHM increases.  
This is consistent with the general relation  
between the spin value of the model and IMEC behavior. As demonstrated in  Ref. \cite{ref36}, when the  spin value of the model increases, the height of the peak in IMEC decreases, but the curves get wider, i.e. FWHM increases.

Besides for lower values of $r$, the double peak behavior of the curve takes attention (see Figs. \re{sek2}
(a) and (b)). This double peak behavior is depressed when the interaction of the core-shell gets stronger
(see Figs. \re{sek2} (c) and (d)). Very recently, this behavior is obtained for the bilayer system experimentally 
\cite{ref33}. Besides, double peak behavior has been obtained theoretically for bilayer \cite{ref34} and
superlattice systems \cite{ref35}.


The same double peak behavior can be seen for the system constituted by spins $S_c=7/2$ (Fig. \re{sek3}). 
But the evolution of the curves by changing shell spin value is slightly different from the 
nanotubes that have $S_c=1/2$, for the nanotubes that have $S_c=7/2$ as a core spin value 
(see Fig. \re{sek3}). When the 
shell spin value increases, the peak value of IMEC increases.

For non-interacting core-shell, two peaked behavior of IMEC occurs, as seen in Figs.   \re{sek2} (a) and  
\re{sek3} (a). For non-interacting case, the system consists of two independent layer which have spin 
values $S_c$ and $S_s$. The low temperature peak seen in   Fig.\re{sek2} (a) is related to the system with spin
value of $S_c$ and other peak is related to the system with spin value $S_s$. Since $S_s>S_c$ in Fig. \re{sek2} (a), 
it is natural for the peak related to the $S_s$ to lie right side of the 
peak related to $S_c$ in ($|\Delta S_M|,t$) plane,
due to the relations between the critical temperatures of layers that have different spin values.The same 
reasoning holds also for  Fig.\re{sek3} (a). When the interaction between the core and shell increases, one peak 
behavior takes place (compare  Figs.\re{sek2} (d) by (a), and Figs.\re{sek3} (d) by (a)). While this transition, 
the peak that occurs at lower temperature values suppressed 
(compare  Figs.\re{sek2} (b) by (a), and Figs.\re{sek3} (b) by (a)).

\begin{figure}[h]\begin{center}
\epsfig{file=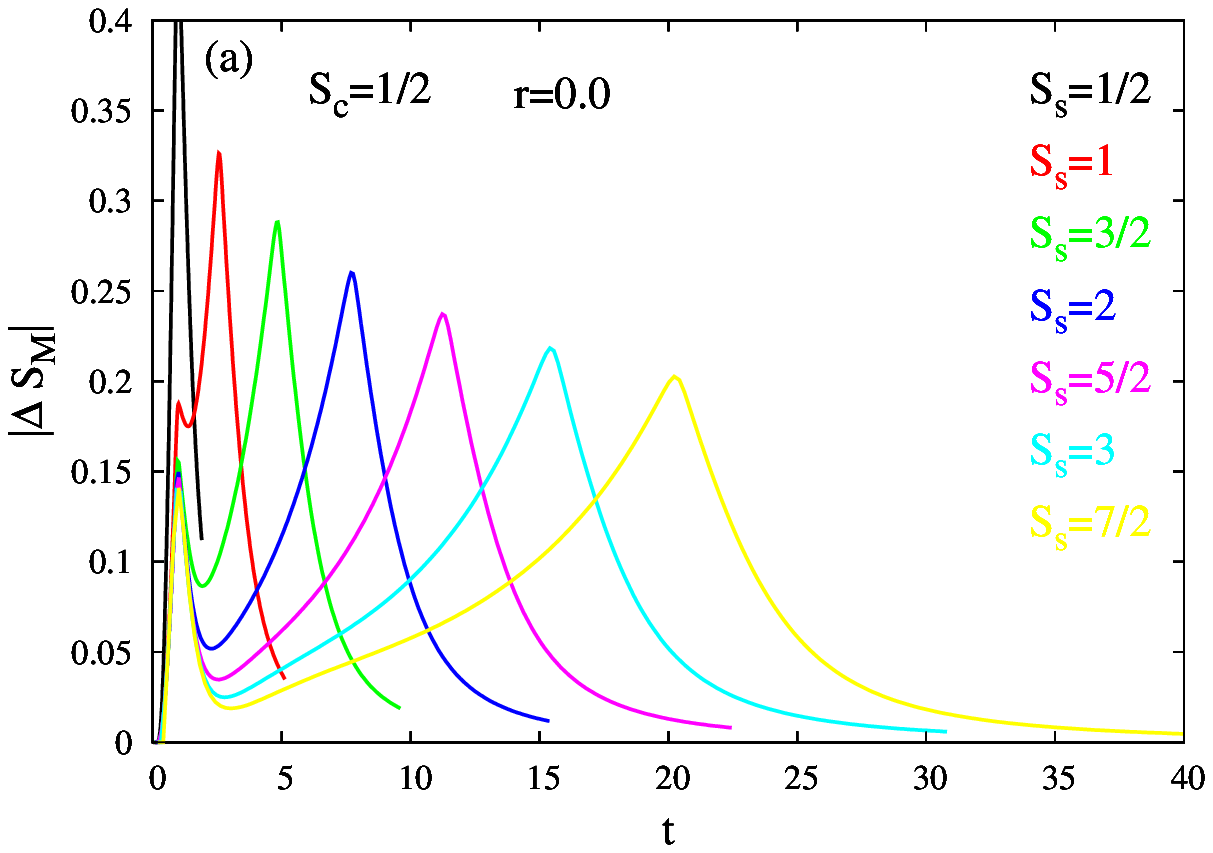, width=6.0cm}
\epsfig{file=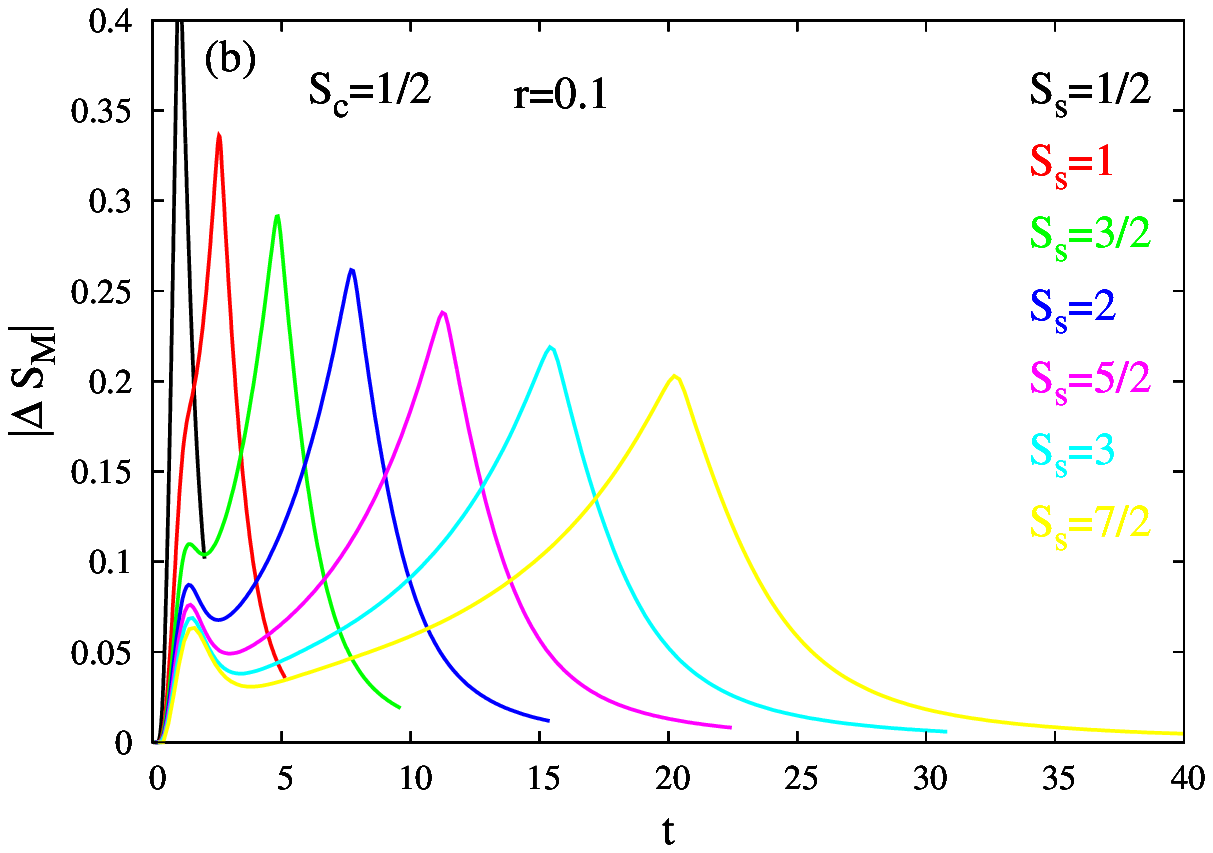, width=6.0cm}
\epsfig{file=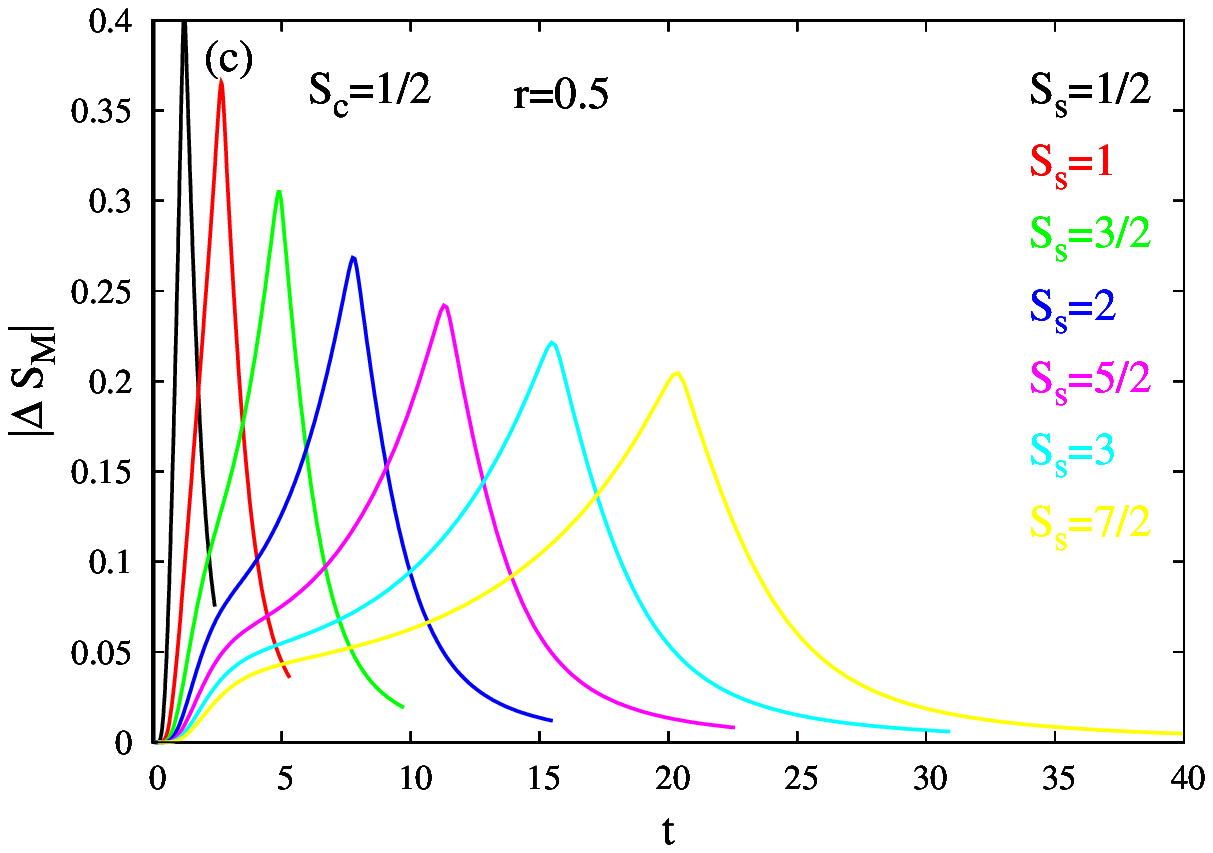, width=6.0cm}
\epsfig{file=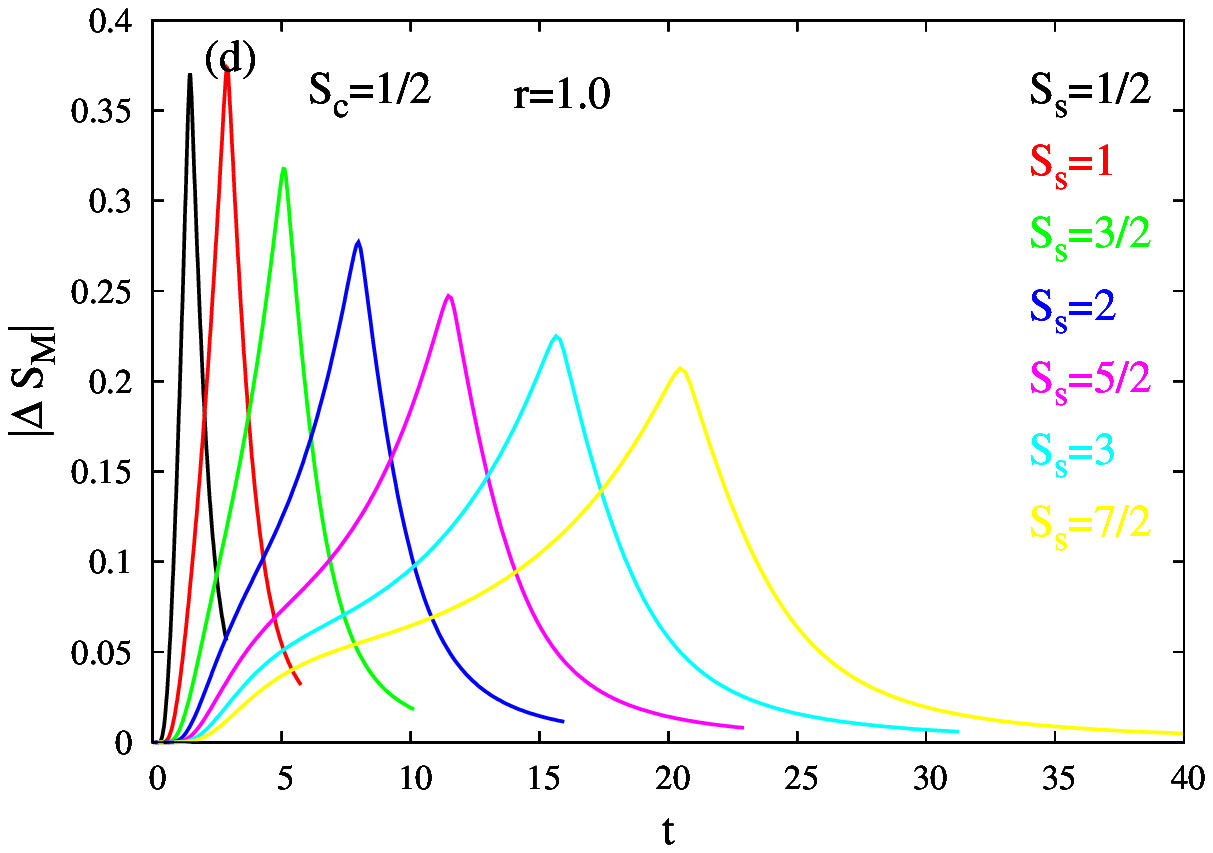, width=6.0cm}
\end{center}
\caption{The variation of IMEC with the temperature for selected values of $S_s=1/2,1,3/2,2,5/2,3,7/2$ and
$r=0.0,0.1,0.5,1.0$ for nanotube that have core spin value of $S_c=1/2$.} \label{sek2}
\end{figure}

\begin{figure}[h]\begin{center}
\epsfig{file=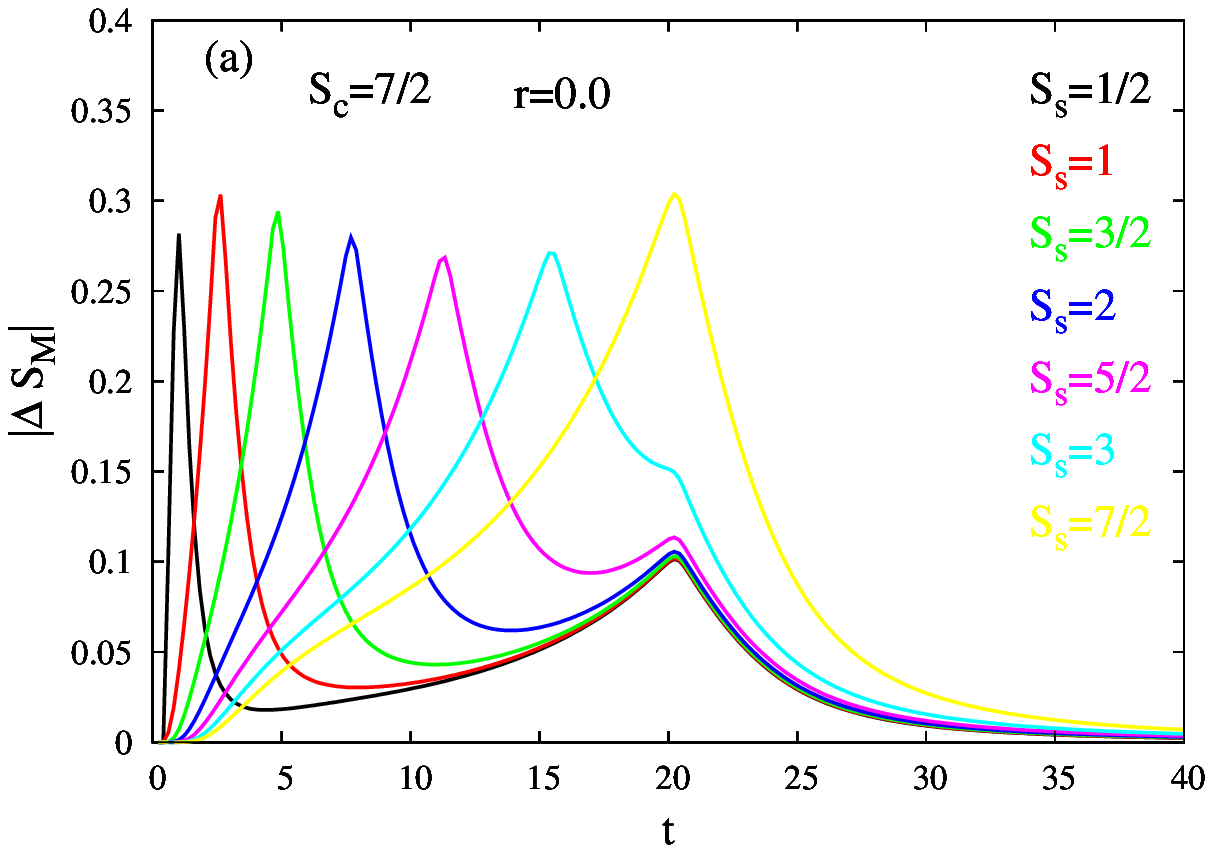, width=6.0cm}
\epsfig{file=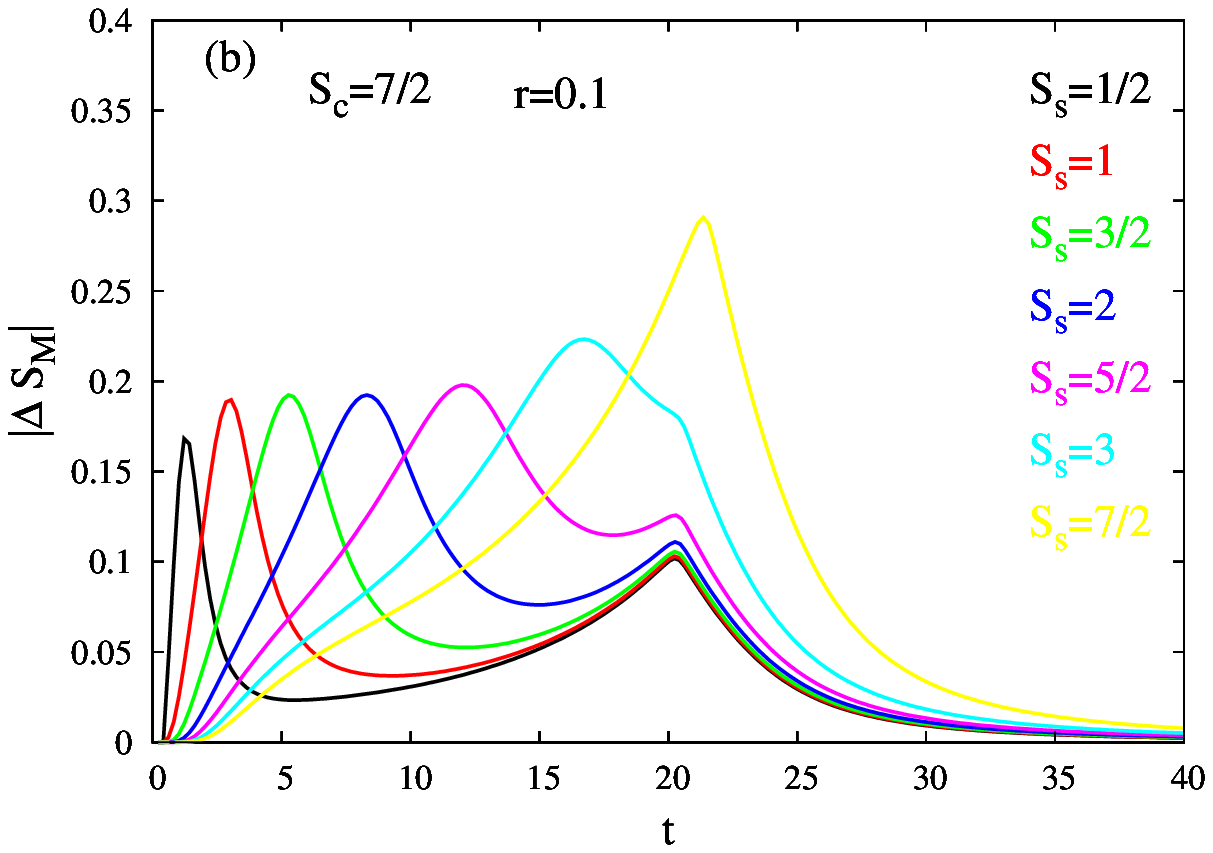, width=6.0cm}
\epsfig{file=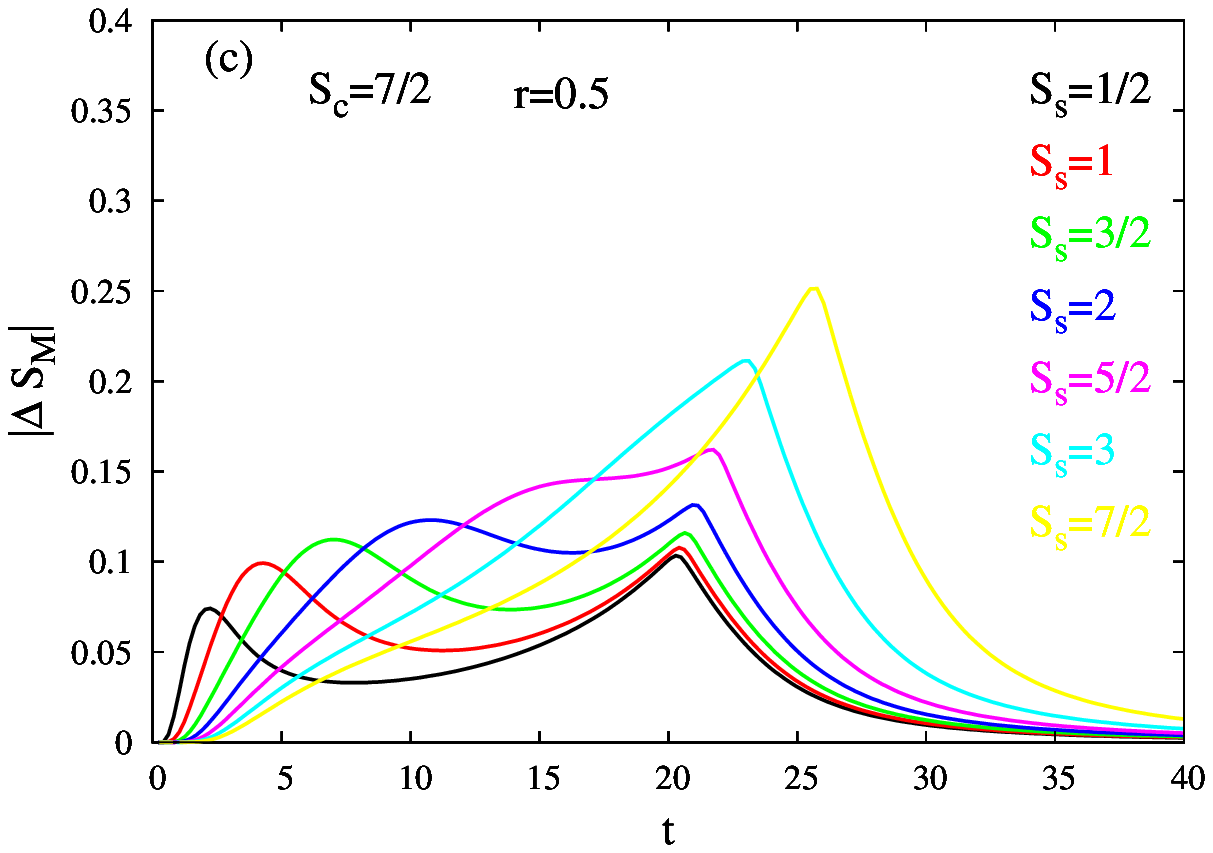, width=6.0cm}
\epsfig{file=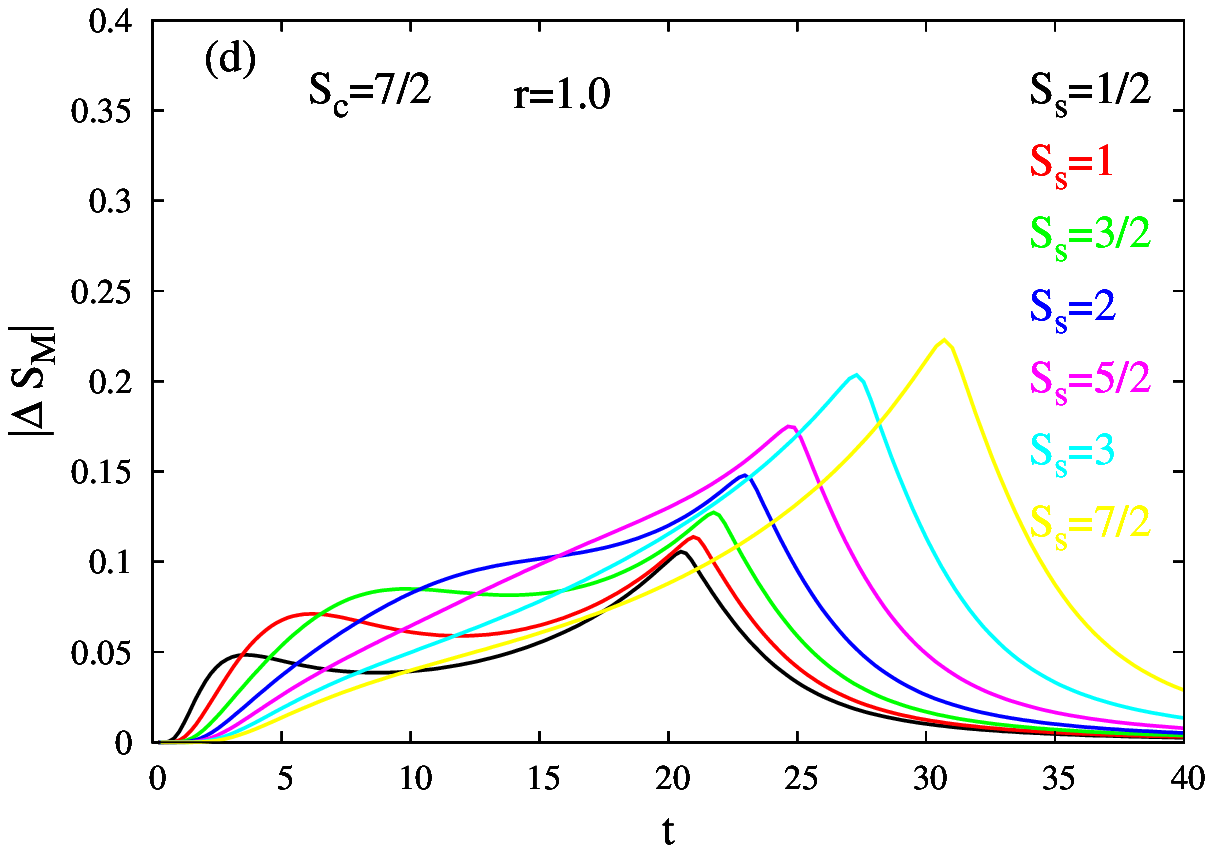, width=6.0cm}
\end{center}
\caption{The variation of IMEC with the temperature for selected values of $S_s=1/2,1,3/2,2,5/2,3,7/2$ and
$r=0.0,0.1,0.5,1.0$ for nanotube that have core spin value of $S_c=7/2$.} \label{sek3}
\end{figure}

To take a closer look at the IMEC behaviors with the spin value and the value of core-shell interaction, we
calculate the maximum value (height of the peak) of the IMEC for different nanotubes which can be seen in Fig. \re{sek4}. 
At first sight, height of the peak of IMEC for a certain $S_c$ occurs at $S_s=S_c$ regardless of the value of
$r$. Thus, as seen in Fig. \re{sek4} decreasing trend with rising $S_s$ occurs for $S_c=1/2$ and increasing 
trend with rising $S_s$ occurs for $S_c=7/2$. For the values of $1/2<S_c<7/2$, rising $S_s$ rises the 
height of the peak of IMEC until $S_s=S_c$, after then rising $S_s$ causes to a decline in  
the height of the peak of IMEC. We can see similar behavior for FWHM in Fig. \re{sek5}. Except  $(S_c,S_s)=(5/2,1/2),
(3,1/2),(7/2,1/2)$ nanotubes, rising $S_s$ first decreases FWHM, minimum FWHM occurs at $S_s=S_c$, after then
rising $S_s$ causes to increment behavior in FWHM. 

\begin{figure}[h]\begin{center}
\epsfig{file=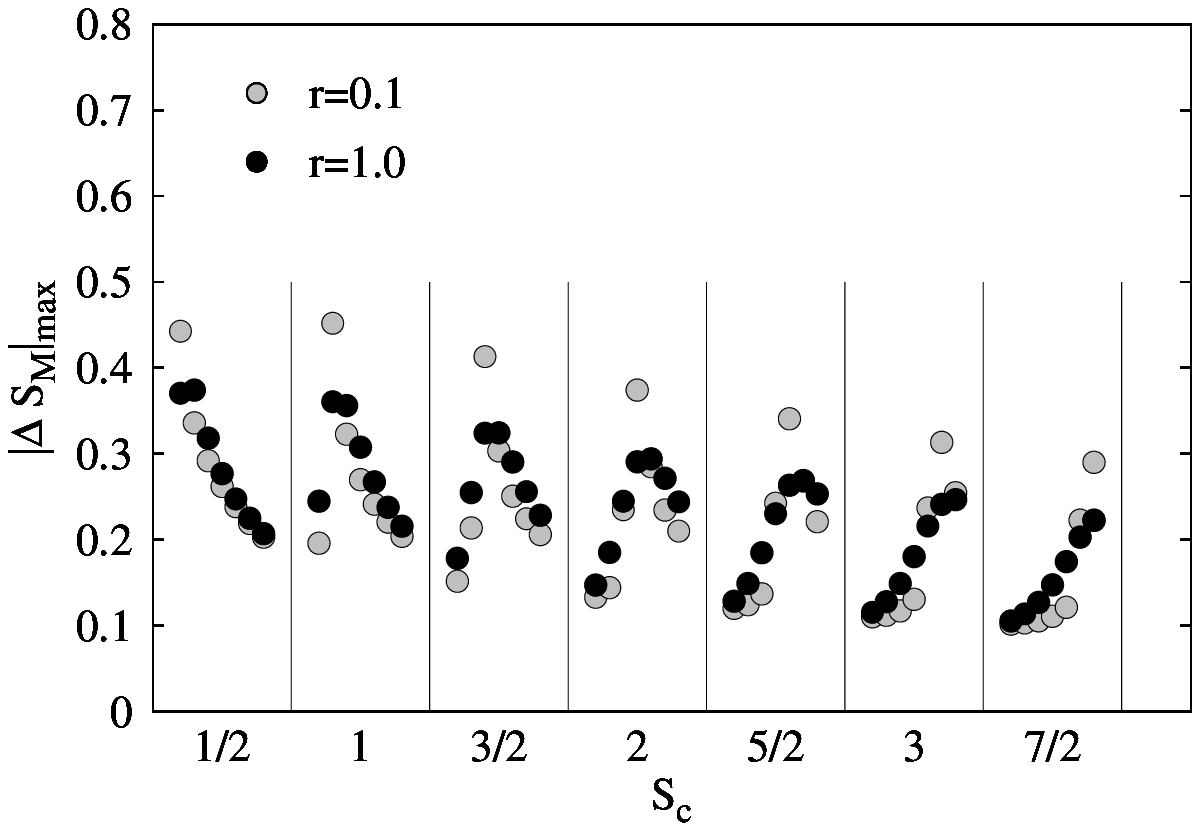, width=12.0cm}
\end{center}
\caption{The maximum value of the IMEC for nanotubes that consist of spin values $S_c,S_s=1/2,1,3/2,2,5/2,3,7/2$
and for selected values of $r=0.1,1.0$. Each box labeled by $S_c$ contains number of 7 circles 
for certain value of $r$. Each circle corresponds to different values of $S_s$, starting from $S_s=1/2$ (most left), 
by increment value of $1/2$ and arrive $S_s=7/2$ (most right) in a box.}
\label{sek4}
\end{figure}

\begin{figure}[h]\begin{center}
\epsfig{file=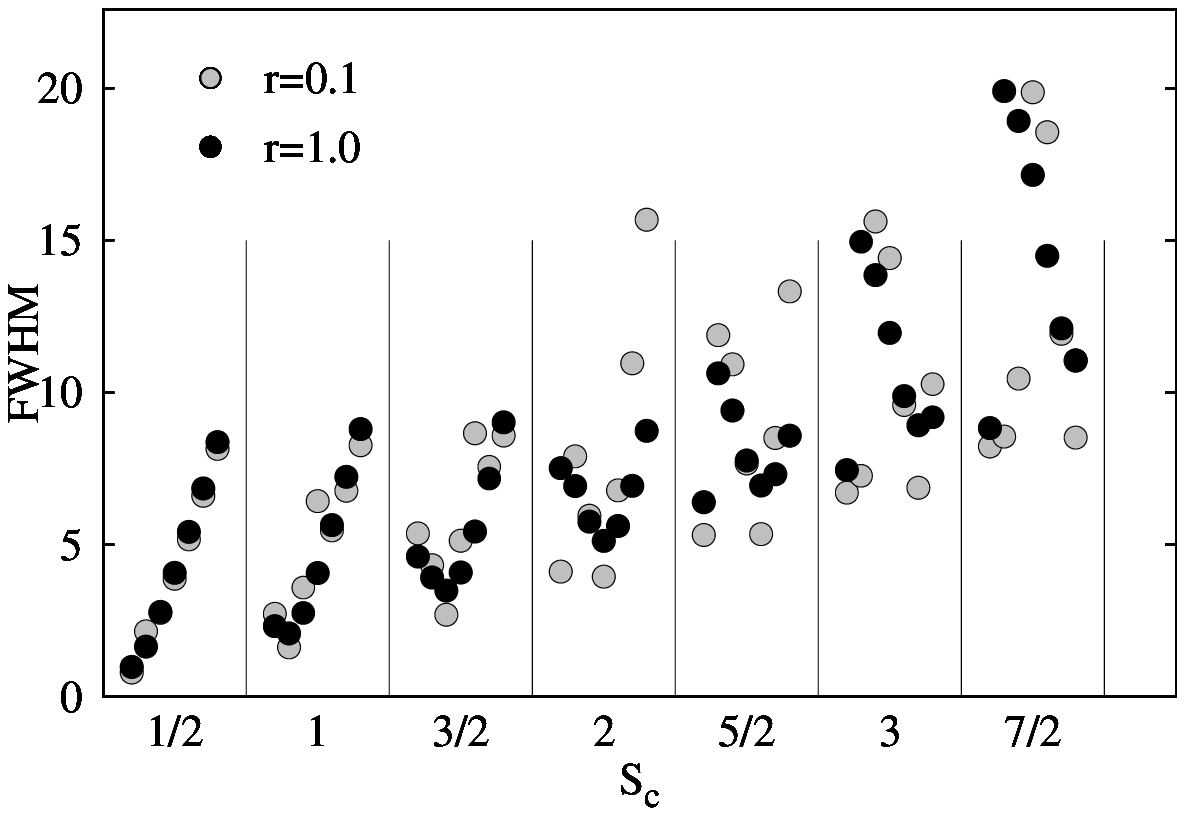, width=12.0cm}
\end{center}
\caption{The value of FWHM for nanotubes that consist of spin values $S_c,S_s=1/2,1,3/2,2,5/2,3,7/2$
and for selected values of $r=0.1,1.0$. Each box labeled by $S_c$ contains number of 7 circles 
for certain value of $r$. Each circle corresponds to different values of $S_s$, starting from $S_s=1/2$ (most left), 
by increment value of $1/2$ and arrive $S_s=7/2$ (most right) in a box.} \label{sek5}
\end{figure}

For refrigerant capacity defined in Eq. \re{denk13}, we depict the same scatter plot in Fig. \re{sek6}. As we can see 
from Fig. \re{sek6}, rising $S_s$ cause increasing refrigerant capacity for spin values  of core provide $S_c<3$. 
If the core spin value exceeds $5/2$, slightly lowering behavior takes place for larger shell spin values.
Interestingly, weak interaction between the core and the shell causes larger refrigerant capacity for higher 
spin values (compare gray dots by black dots in $S_c=3$ and $7/2$).

\begin{figure}[h]\begin{center}
\epsfig{file=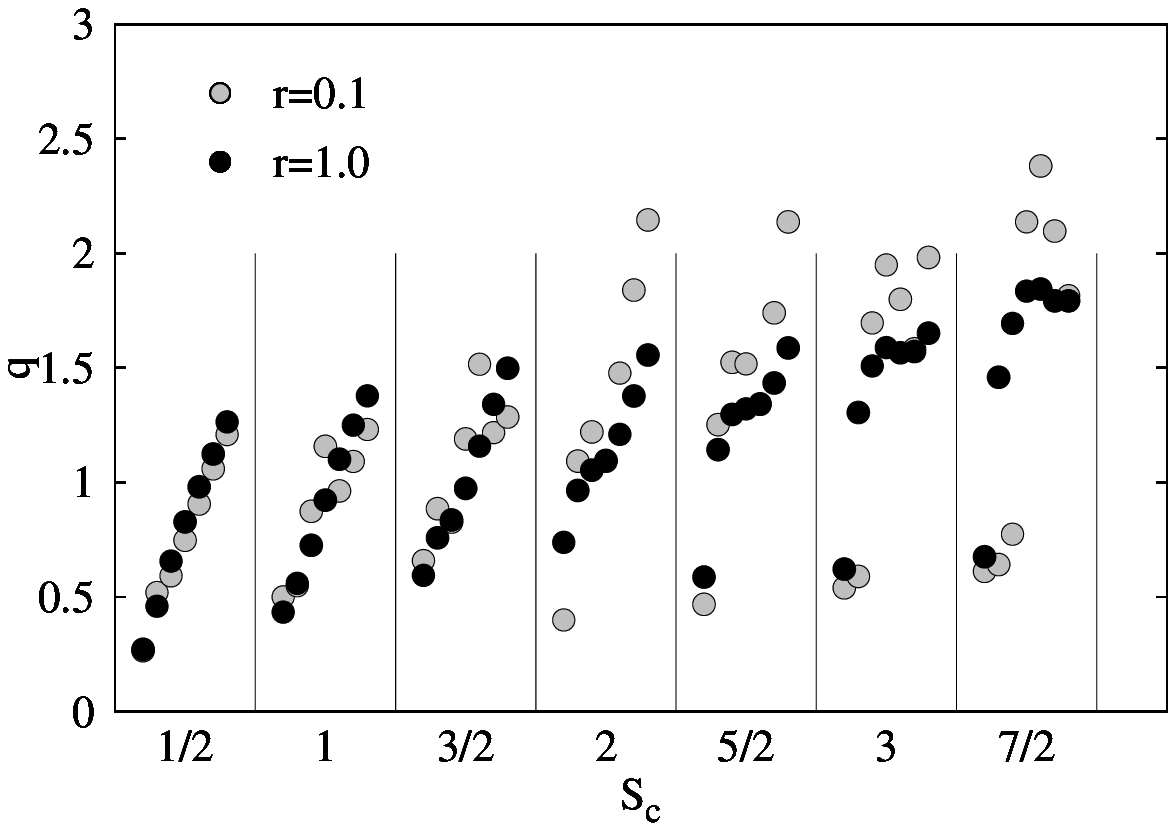, width=12.0cm}
\end{center}
\caption{The value of the refrigerant capacity ($q$) for nanotubes that consist of spin values $S_c,S_s=1/2,1,3/2,2,5/2,3,7/2$
and for selected values of $r=0.1,1.0$. Each box labeled by $S_c$ contains number of 7 circles 
for certain value of $r$. Each circle corresponds to different values of $S_s$, starting from $S_s=1/2$ (most left), 
by increment value of $1/2$ and arrive $S_s=7/2$ (most right) in a box.} \label{sek6}
\end{figure}

\section{Conclusion}\label{conclusion}

The MCE properties of the Ising nanotube constituted by arbitrary core spin 
values $S_c$ and the shell spin values $S_s$ have been investigated by 
mean field approximation. During this investigation, several quantities 
have been calculated, such as IMEC, FWHM and the refrigerant capacity ($q$). 
The variation of these quantities with the values of the spins and exchange interaction
between the core and shell is determined. 

First general conclusions about the variation of the IMEC with the
temperature has been obtained. As consistently by the conclusions obtained in Ref. \cite{ref36} 
for the general spin valued Ising model on a regular lattice, it is observed that when 
the spin values of the nanotube increase, the height of the peak in IMEC decreases. This peak occurs at the critical temperature of the system, as expected. Besides, 
for a chosen spin value for the core, increasing shell spin value yields rising height of the peak in IMEC, when $S_c=S_s$ maximum value is obtained. After that (i.e. $S_c<S_s$), increasing spin value of the shell yields decreasing behavior in  the height of the peak in IMEC. Completely reverse evolution occurs in FWHM, when the value of the shell spin increases. On the other hand, refrigerant capacity has increasing trend in the conditions of rising core and shell spin values. These observations may yield a tuning of MCE in nanotube system. Although it is very hard task to tune the interaction between the core and shell experimentally, theoretical knowledge about the relation between the spin values (or exchange interaction between the core and the shell) and MCE characteristics may yield important experimental achievements.

Other than these results, recently obtained double peak behavior in IMEC for bilayer system is observed in nanotube system also. The physical explanation is briefly discussed.

We hope that the results  obtained in this work may be beneficial form both 
theoretical and experimental point of view.

\end{document}